\documentclass[a4paper,11pt]{article}
\pdfoutput=1 % if your are submitting a pdflatex (i.e. if you have
\usepackage{jinstpub} % for details on the use of the package, please
\usepackage{siunitx}
\usepackage{chemformula}
\usepackage{color}
\usepackage{ulem}
\title{\boldmath Suppression of electrical breakdown phenomena in liquid TriMethyl Bismuth based ionization detectors}

\author[1]{Bj\"{o}rn Gerke,\note{Corresponding author.}}
\author[2]{Simon-Nis Peters,}
\author[2]{Nils Marquardt,}
\author[2]{Christian Huhmann,}
\author[2]{Volker Michael Hannen,}
\author[3]{Michael Holtkamp,}
\author[3]{Uwe Karst,}
\author[4,5]{Dominique Yvon,}
\author[4,5]{Viatcheslav Sharyy,}
\author[2]{Christian Weinheimer,}
\author[1]{Klaus Sch\"{a}fers}

\affiliation[1]{European Institute for Molecular Imaging, University of M\"{u}nster, M\"{u}nster, Germany}
\affiliation[2]{Institute for Nuclear Physics, University of M\"{u}nster, M\"{u}nster, Germany}
\affiliation[3]{Institute of Inorganic and Analytical Chemistry, University of M\"{u}nster, M\"{u}nster, Germany}
\affiliation[4]{IRFU, CEA, Universit\'{e} Paris-Saclay, Gif-sur-Yvette, France}
\affiliation[5]{BioMAPs, Service Hospitalier Fr\'{e}d\'{e}ric Joliot, CEA, CNRS, Inserm, Universit\'{e} Paris-Saclay, Orsay, France}

\emailAdd{gerkeb@uni-muenster.de}

\abstract{

Organometallic liquids provide good properties for ionization detectors. TriMethyl Bismuth (TMBi) has been proposed as a detector medium with charge and Cherenkov photon readout for Positron Emission Tomography. 
In this work, we present studies for the handling of TMBi at different electric fields and under different environmental conditions to find applicable configurations for the suppression of electrical breakdowns in TMBi at room temperature. A simple glass cell with two electrodes filled with TMBi was constructed and tested under different operation conditions.
Working at the vapour pressure of TMBi at room temperature of about 40~mbar and electric fields of up to $\SI{20}{kV/cm}$ in presence of a small oxygen contamination we found the formation of a discharge channel in the liquid and a steady increase in the current. Further reduction of pressure by pumping caused the TMBi to boil and a spontaneous combustion. Eliminating the oxygen contamination led the TMBi under the same condition to only decompose.
When operating the setup under an argon atmosphere of 1~bar we did not observe breakdowns of the electrical potential up to field strengths of $\SI{20}{kV/cm}$. Still, in presence of a small oxygen contamination fluctuating currents in the nA range were observed, but no decomposition or combustion.
We conclude from our experiments that TMBi at room temperature in a pure argon atmosphere of 1~bar remains stable against electrical breakdown at least up to electric field strengths of $\SI{20}{kV/cm}$, presumably because the formation of gaseous TMBi was prevented.} 


\proceeding{European Institute for Molecular Imaging\\
  June 2022\\
  University of Muenster}

\begin{document}
\maketitle

\flushbottom

\section{Introduction}
Liquid ionization detectors have been proposed for many applications both in low energy particle physics as well as medical imaging \cite{klages1997kascade}\cite{xing2017xemis}\cite{manzano2018xemis2}. Noble element liquid detectors, e.g. based on argon or xenon, offer a promising combination of excellent scintillation properties in conjunction with the capability to drift charges \cite{aprile2010liquid}. They find application in large-scale dark matter \cite{aprile2020projected}\cite{aprile2021search}\cite{meng2021dark}\cite{akerib2020lux}\cite{aalbers2016darwin} and neutrinoless double beta decay experiments \cite{adhikari2021nexo}\cite{adams2021sensitivity}.
While liquid xenon is a very effective detector material with high density (\SI{5.9}{g/\cubic\cm}) and high Z (54), cryogenic temperatures are required for operation, as has been reported for its use in a time projection chamber (\SI{-96}{\degreeCelsius} at \SI{1.935}{bar}) \cite{aprile2017xenon1t}. This is of some limitation for such a detector system if it would be used in a medical environment. 
Nevertheless, new detector concepts for positron emission tomography (PET) have been introduced recently based on liquid xenon detectors with the advantage of providing large area detectors \cite{manzano2015xemis}\cite{ferrario2022status}. This is of special interest in whole-body sized PET scanners for simplifying the detector setup which is commonly built on thousands of small pixelated detector elements.

With the idea of operating at room temperature, a new drift detector has been proposed for PET based on an heavy organometallic liquid - TriMethyl Bismuth (TMBi) \cite{Yvon2014}\cite{farradeche2020ionization}. TMBi (chemical formula \ch{(CH3)3Bi}) is a transparent dielectric liquid consisting of three methyl groups and one bismuth atom. Because of the high atomic number of bismuth (Z\textsubscript{Bi}~=~83) TMBi effectively converts \SI{511}{keV} photons by the photo-electric effect ($47~\%$ photo fraction) offering the possibility to read out both charges in a high resolution drift detector as well as Cherenkov light of the primary photo-electron for fast event timing. High electric field strengths are required to separate electron-ion pairs generated in the liquid and to drift the electrons towards a segmented anode.
The liquid must be highly purified to reduce recombination \cite{freeman1973onsager}\cite{onsager1938initial}. However, high field strengths can trigger an electric breakdown in the liquid generating high local temperatures. Temperatures above 106~\textdegree C lead to a decomposition of 
TMBi resulting in a sudden increase in volume. Such an event happened recently at \textit{IRFU, CEA, Universit\'{e} Paris-Saclay} causing an unexpected explosion during an experiment \cite{farradeche}.

In this work we present test measurements regarding the handling of TMBi at different electric fields and under different environmental conditions in order to find useful configurations for the suppression of electrical breakdowns in TMBi.

\label{sec:model}

\section{Electrical breakdown phenomena}
The following sections discuss mechanisms of electric breakdown in gases and dielectric liquids.

\subsection{Electrical breakdown in gases}

\begin{figure}[ht!]
\centering
\includegraphics[width=0.95\columnwidth]{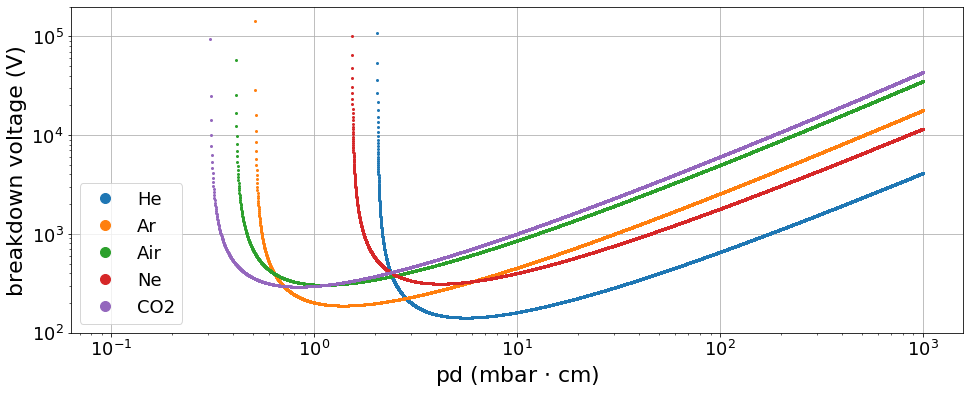}
\caption{Paschen curves for different gases over a wide range of values for $pd$, calculated for stainless steel electrodes ($\gamma = 0.01$).}
\label{fig:paschen06}
\end{figure}

Electrical breakdown in gases has significance for many technical applications like the glow discharge in fluorescent tubes, electric arcs in welding or avalanches of charge carriers in a Geiger-Müller detector. In drift detector applications, electrons and ions generated in a neutral gas volume are accelerated by an electric field. As the electric field strength increases, recombination of charge carriers is reduced and more electrons reach the anode contributing to an electric current until all generated charges are collected. 
When the electrons are accelerated to a higher kinetic energy than the ionization energy of the gas molecules, secondary charge carriers are created, resulting in an avalanche of charged particles. This effect, referred to as Townsend avalanche \cite{townsend1900conductivity}, may trigger an electrical breakdown which is utilized in Geiger-Müller counters but must be avoided for the operation of a proportional drift detector.
The Townsend discharge can be described by Paschen's law (Eq. \ref{eqn:paschen}), where the breakdown voltage $U\textsubscript{b}$ depends primarily on the product ($pd$) of pressure and distance between two parallel electrodes \cite{pejovic2002electrical}.
Without supply of external electrons, i.e. without radioactive source, and for parallel plane electrodes the breakdown voltage can be calculated using
\begin{align}
\label{eqn:paschen}
U\textsubscript{b} &= \dfrac{B \cdot pd}{\ln (A \cdot pd) - \ln\left(\ln \left(1 + \dfrac{1}{\gamma}\right)\right)},
\end{align}
where $\gamma$ represents the second Townsend coefficient ($\gamma=0.01$ for stainless steel electrodes) and A and B are empirically determined gas coefficients (see table \ref{table:coefficient}) \cite{burm2007calculation} \cite{von1965ionized} \cite{norman2022dielectric} \cite{Massarczyk_2017}.
\begin{table}[htbp]
\centering
\caption{\label{tab:i1} Constants for the Townsend ionization coefficients based on values from \cite{raizer1991gas}.}
\smallskip
\begin{tabular}{|c|c|c|}
\hline
Gas & A ($\SI{}{mbar\textsuperscript{-1}~\cdot~cm\textsuperscript{-1}}$) & B ($\SI{}{V~/~(mbar~\cdot~cm)}$) \\
\hline
He & 2 & 26 \\
Ar & 9 & 135 \\
Air & 11 & 274 \\
Ne & 3 & 751\\
\ch{CO2} & 15 & 350\\
\hline
\end{tabular}
\label{table:coefficient}
\end{table}
For the gases He, Ar, Air, Ne and \ch{CO2}, the break down voltages reach a minimum within a pd range between 0.66\;\SI{}{mbar\cdot cm} and 6\;\SI{}{mbar\cdot cm} (see figure~\ref{fig:paschen06}).
Left of the Paschen's minima the probability of an avalanche effect is low due to a lack of gas molecules. At higher $pd$ values, the breakdown voltage increases providing stable operating conditions when used in drift detectors. At low pressure near the Paschen's minima a glow discharge may form when charge carriers obtain enough kinetic energy from the electric field to ionize further gas molecules. 
In an arc discharge it is essentially the emission of electrons from the cathode that contributes to the creation of the DC current. The electric current is much higher than in the case of a glow discharge and thus requires a smaller resistance of the external DC circuit \cite{raizer1991gas} \cite{kurt2003behaviour} \cite{lisovskii2000modified}.

\subsection {Electrical breakdown in dielectric liquids}
\label{sec:breakdown}
The occurrence of electrical breakdowns in dielectric liquids depends on many factors such as the chemical composition of the liquid and level of impurities, temperature and pressure, irradiation and applied voltages, as well as electrode geometry and surface properties. In the following, only phenomena for DC circuits are considered which are relevant to our experiments \cite{schmidt1997liquid}. Two common theories exist to describe the initiation of electrical breakdown in dielectric fluids, namely the bubble theory and the direct impact ionization theory \cite{Wedin2014}.
The bubble theory assumes that electrical breakdown begins in a gas bubble near or at an electrode~\cite{Sharbaugh1978}. The generation of such a bubble is the result of an energy injection causing an electron avalanche in the liquid. The formation and size of the bubbles ($\approx \SI{}{\micro\m}$) and the live time ($\approx \SI{}{\micro\s}$) depends on the operating pressure The bubbles deform due to Coulomb forces in the direction of the electric field and weaken the liquid's breakdown strength. The elongation of the bubbles leads to an increase in the electric field near the bubble poles. When the electric field reaches a critical value, electrical breakdown starts in the fluid. A Townsend discharge in the bubble can introduce additional electrons in the liquid at the bubble poles, further promoting a breakdown \cite{korobeynikov1998bubbles}\cite{alexeff1990possible}\cite{jones1995development}.
With increasing pressure, the bubble volume decreases and the dielectric strength increases \cite{kattan1989generation}. \\
The theory of direct impact ionization provides a description of breakdown effects in liquids without initial bubbles and can be classified into different phases \cite{andre2011conduction}:
\begin{enumerate}
 \item A trigger phase before breakdown. High electric fields above a few $\SI{100}{kV/mm}$, which might occur at micro tips on charged components in the setup, cause local discharges. This will lead to the formation of cavities or volumes with reduced density. The resulting structures form channels in the liquid with a different density than the surrounding medium which in turn leads to an increase in electric current. 
 \item A phase of expansion of a channel in the liquid. It is called a streamer, which is further divided into positive and negative streamers. Lower electric fields on the magnitude of a few $\SI{1}{kV/mm}$ are required for generation.
 \item A phase of transition to an arc discharge after the streamer short-circuits the electrodes and produces a high current in a low-impedance circuit.
\end{enumerate}
Positive streamers spread out in the direction of the electric field and against the direction of movement of the drift electrons. During the spread the current growths continuously leading to a breakdown quickly after the streamer reaches the cathode \cite{Devins1981}. Positive streamers require lower electric fields and are more filamentary than negative streamers and pose a higher risk for technical applications \cite{lesaint1994gaseous} \cite{Tobazcon}. 
Negative streamers spread out against the direction of the electric field and in the direction of movement of the drift electrons. They are less filamentary than positive streamers \cite{Devins1981}. The current during growth consists of successive fast pulses. In contrast to positive streamers, the influence of pressure on the spread of negative streamers is stronger \cite{beroual1998propagation} \cite{denat2006high}.\\
The influence of pressure on dielectric strength is evident both in the formation of bubbles and in the initialization of direct impact ionization. With increasing pressure, the dielectric strength increases. Correspondingly, the number and amplitude of the current pulses, as well as the shape and length of the streamers, decrease \cite{Devins1981} \cite{beroual1998propagation} \cite{Watson_1960} \cite{beroual1985effects}. A minimum dielectric strength is reached near the boiling point of the liquid or near the boiling points of residual gas inclusions \cite{CLARK1933429} \cite{Koo1961}. In studies with point-plane geometries, the transition to a streamer phenomenon is observed as a result of fast and repeated energy injection by electron avalanches, which correlates with the generation of successive bubbles \cite{kattan1989generation} . 
The dielectric strength depends also on the nature and quantity of impurities in the liquid~\cite{Devins1981}. Even small contaminations can lead to an increased current and a reduction in breakdown voltage. In tests with liquid argon, the breakdown voltage was a factor $\approx 1.5$ higher for levels of oxygen impurities above $\SI{0.2}{ppm}$ than for concentrations below $\SI{1.8}{ppb}$~\cite{Acciarri_2014}.\\
Electrical breakdowns must be avoided when operating a drift detector and even more importantly when using a pyrophoric liquid such as TMBi. 
In the experiments described in this paper, conditions that favor or prevent electrical breakdowns in TMBi have been investigated and methods for stable operation are discussed.

\subsection{Electrical breakdown behavior of TMBi}
\begin{figure}[ht!]
\centering
\includegraphics[width=1\columnwidth]{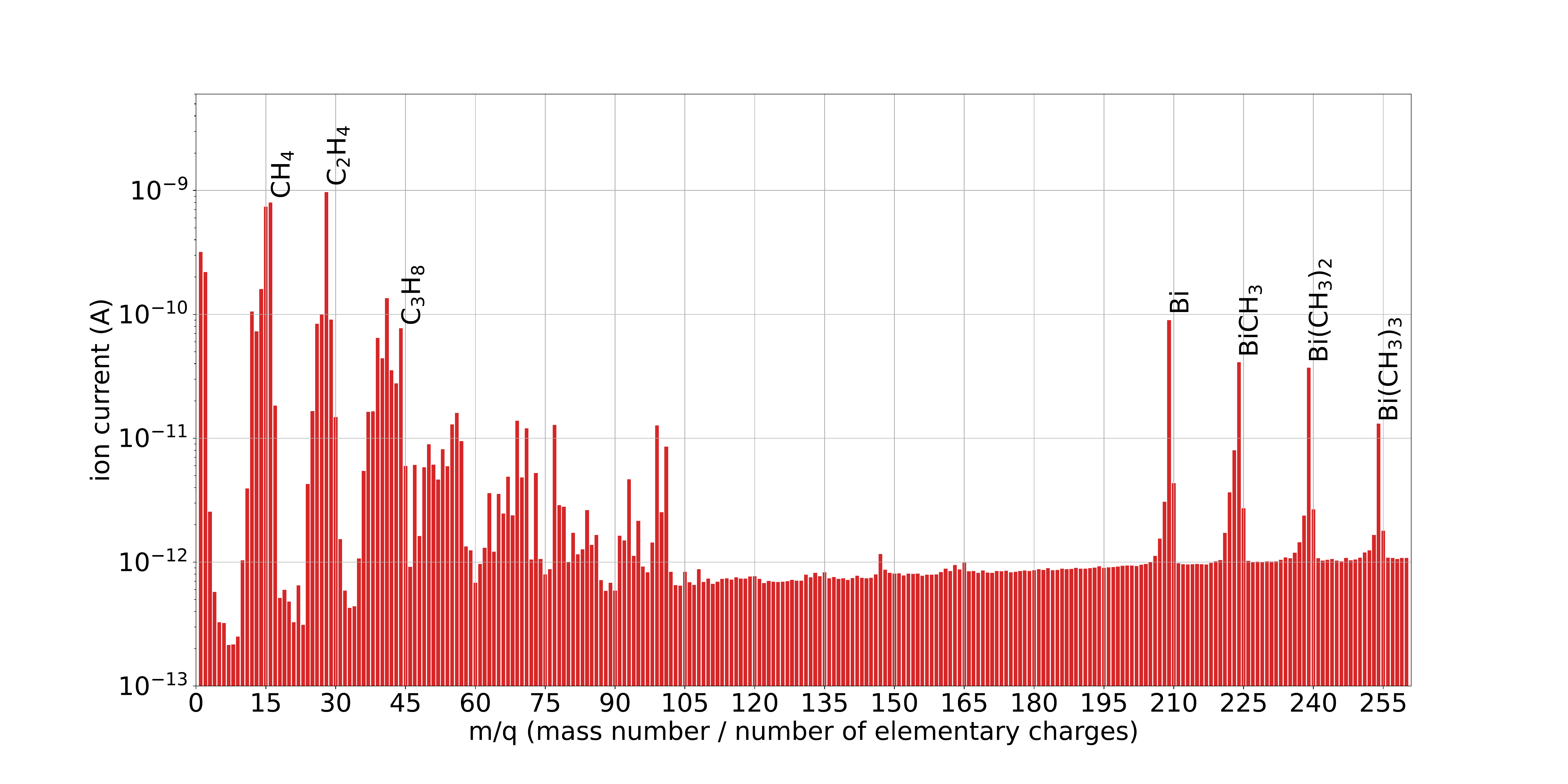}
\caption{Spectrum of decomposition products of trimethyl bismuth taken with a quadrupole mass filter. The main components at mass numbers above 200 are Bi (\SI{209}{amu}), \ch{Bi(CH3)} (\SI{224}{amu}), \ch{Bi(CH3)2} (\SI{239}{amu}), and \ch{Bi(CH3)3} (\SI{254}{amu}). The various 
hydrocarbon compounds like ethene \ch{C2H4} (\SI{28}{amu}), ethane \ch{C2H6} (\SI{30}{amu}), and propane \ch{C3H8} (\SI{44}{amu}) appear at lower mass numbers.}
\label{fig:RGA}
\end{figure}

In an electrical breakdown (in the gas phase), TMBi decomposes via an exothermic redox reaction into lower methyl bismuth species including \ch{(CH3)Bi} and \ch{(CH3)2Bi} \cite{strausz1971flash}. Similar decomposition products are generated during the electron bombardment in the ion source of a quadrupole mass spectrometer (PrismaPro QMG 250 M3, Pfeiffer Vacuum GmbH, Germany) (see figure \ref{fig:RGA}). The difference between the bond enthalpies of the products and reactants is converted into heat. If the temperature becomes sufficiently high, TMBi decomposes in a chain reaction \cite{lide2004crc}\cite{riedel2022anorganische}. If oxygen comes into contact with the TMBi, carbon, or hydrocarbon residues during the redox reaction, this can lead to combustion (e.g. into $\mathrm{Bi_2O_3}$, $\mathrm{CO_2}$ and water) with the formation of flames \cite{long1954heat}. Both effects, decomposition and combustion, will lead to a sudden volume expansion and therefore present a safety hazard.

\label{sec:electrical1}

\section{Setup for breakdown experiments}
\begin{figure}[h]
 \centering
 \includegraphics[width=1\columnwidth]{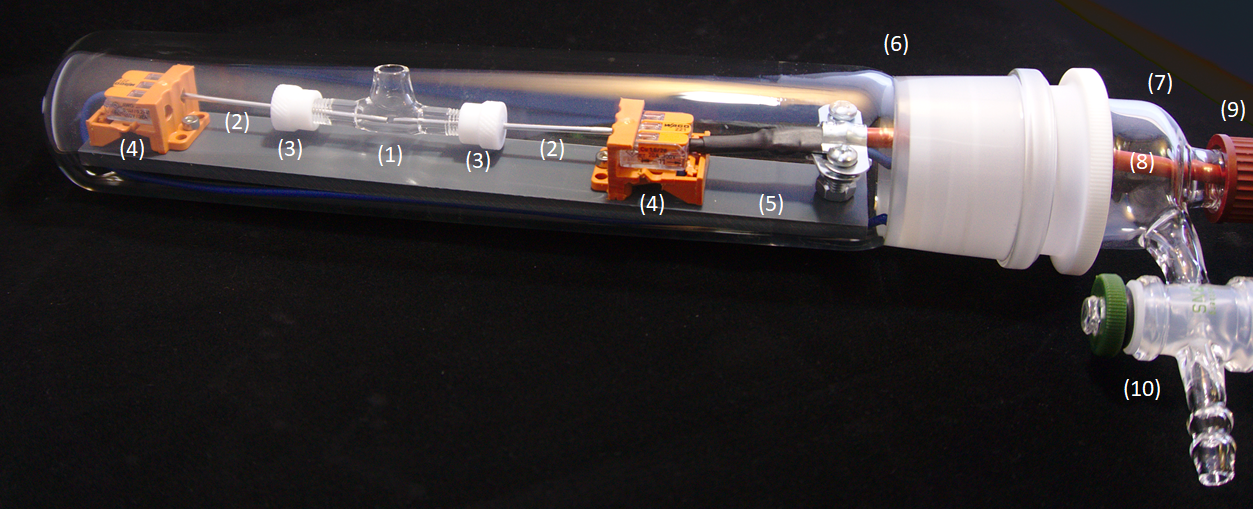}
 \caption{Experimental setup with sample glass container (1) in which high voltage can be applied to the TMBi liquid by two electropolished stainless steel electrodes (distance of $\SI{2}{mm}$). The sample container is sealed by an outer glass cylinder (6) which can be evacuated using a water jet pump.}
 \label{fig:Tube}
\end{figure}

To study the behaviour of TMBi under the influence of different electric fields and different environmental conditions, the liquid was filled into a small glass vessel with electrodes introduced on both ends and hermetically sealed inside a large glass cylinder (see figure \ref{fig:Tube}). 
The sample vessel (1) was manufactured by our local glassblower with an opening on top and a volume of $V \approx \SI{1.5}{\ml}$. Two stainless steel rods (2) with electropolished front faces and a diameter of $\SI{2}{mm}$ were mounted via PTFE inlets (3) on both ends of the glass vessel, serving as electrodes. They had a distance of $\SI{2}{mm}$ and were contacted using spring-loaded clamps (4). Vessel and contacts were mounted on a plastic base-plate (5) which was arranged in an outer glass cylinder (6) (Lenz Laborglas GmbH \& Co.KG, Wertheim, Germany) with a conical ground joint (45/40) and a head closure (7).
A high-voltage cable (8) was fed through a rubber sealed entry cap fixed with a twist lock (9). A one-way connecting stopcock (10) (Gebr. Rettberg GmbH, Göttingen, Germany) was connected to a water jet pump (BRAND GMBH + CO KG, Wertheim, Germany) to achieve pressures down to $p\approx\SI{20}{mbar}$ inside the cylinder.
High voltage was supplied to the electrodes using an iseg NHQ 224M HV module (iseg Spezialelektronik GmbH, Radeberg, Germany) offering an adjustable voltage up to $\SI{4000}{V}$ and an integrated display of the current with a resolution of $\SI{1}{nA}$. All experiments were performed inside a laboratory fume hood with the setup arranged behind a thick perspex window for safety. 

A high speed video camera (acA2000-165um, Basler AG, Ahrensburg, Germany, max. 165 frames per second) was placed in front of the experiment to monitor the test cell. In order to estimate the pressure generated by the water jet pump, a measurement was carried out beforehand in a vacuum system with a comparable volume to the test set-up. The pressure measured using a dedicated sensor (Pirani Gauge FRG-700, Agilent Technologies, Santa Clara, CA, USA) dropped exponentially with time reaching a minimum of \SI{20}{mbar} after \SI{160}{seconds}.

\label{sec:method}

\section{Measurement results}
The following sections describe a series of experiments using TMBi under different atmospheres and at different ambient pressures. The tests were prepared by filling TMBi into the sample container and sealing it inside the glass cylinder described in section~\ref{sec:method} under a clean argon atmosphere inside a glove box. The closed cylinder was then transferred to the laboratory hood and connected to the water jet pump and the high voltage supply.
The negative high voltage terminal of the power supply was connected to the cathode and the anode was connected to ground potential. In experiments 1~-~3, the seal of the high voltage cable entry into the cylinder turned out to be leaky which allowed to study effects of oxygen contamination at different pressures. 
In experiments 4 to 5, the high-voltage cable feed-through was made airtight and additionally the glass cylinder was placed into a second container filled with argon to prevent entry of oxygen during the evacuation phase.
\subsection{Experiment 1: 1 bar argon atmosphere and oxygen contamination.}
\begin{figure}[h]
 \centering
 \includegraphics[width=0.50\columnwidth]{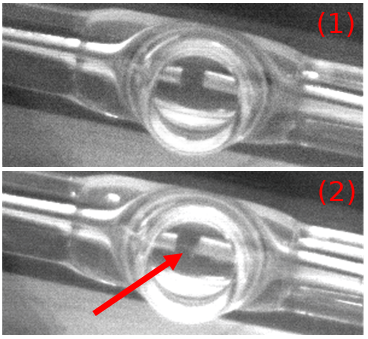}
 \caption{Upper panel: At a pressure of \SI{1}{bar} (experiment 1) no discharges appeared up to $\SI{4000}{V}$. Lower panel: In experiment 2, a discharge channel formed between the electrodes at $\SI{2000}{V}$.}
 \label{fig:TMBi_discharge02}
\end{figure}

In the first test, the TMBi filled vessel was kept inside the argon filled cylinder under atmospheric pressure while applying increasingly negative voltages to the cathode.
The voltage was increased in $\SI{500}{V}$ steps up to $\SI{4000}{V}$ (maximum voltage of the HV supply) in 1 minute intervals.
Starting at a voltage of $\SI{2000}{V}$ visible movements in the liquid with floating particles moving close to the electrodes were observed. The particles are believed to be reaction products of TMBi and oxygen.
After increasing the voltage further to $\SI{3000}{V}$, the measured current started to oscillate between $\SI{0}{nA}$~and~$\SI{3}{nA}$ over time intervals of a few seconds. 
At even higher voltages of $\SI{3500}{V}$ and $\SI{4000}{V}$ the fluctuation amplitudes increased to $\SI{4}{nA}$ and $\SI{6}{nA}$, respectively.
Figure~\ref{fig:TMBi_discharge02} shows on the upper panel a photograph of the TMBi filled sample vessel during the test.

\subsection{Experiment 2: vapor pressure (40~mbar) and oxygen contamination.}
\begin{figure}[ht!]
 \centering
 \includegraphics[width=1 \columnwidth]{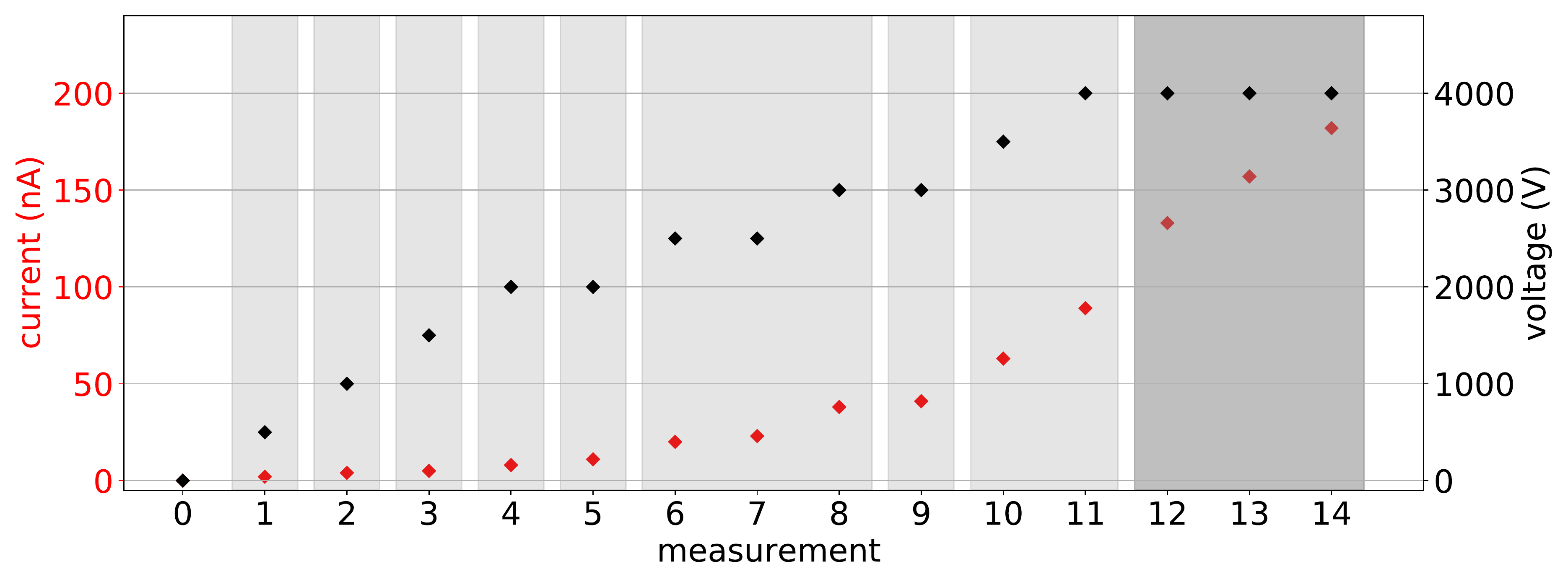}
 \caption{In experiment 2, the current increased as a function of the applied voltage (light grey equals to \SI{1}{min}, dark grey to \SI{10}{min} time scales). At \SI{4000}{V}, the current increased continuously and the test was stopped.}
 \label{fig:TMBi_exp2}
\end{figure}

For the second test, the voltage was turned down and the outer cylinder was evacuated using the water jet pump until the TMBi started to boil. At that point, the valve to the pump was closed and the pressure inside the cylinder increased to \SI{40}{mbar} corresponding to the vapor pressure of TMBi at room temperature \cite{amberger1961hydride}\cite{moravek2013vapor}.
While increasing the voltage from $\SI{0}{V}$ to $\SI{4000}{V}$ in $\SI{500}{V}$ steps at this lower pressure, the current oscillations already started at $\SI{500}{V}$ where we observed fluctuations up to $\SI{2}{nA}$. These fluctuations increased to $\SI{4}{nA}$ at $\SI{1000}{V}$ and up to $\SI{5}{nA}$ at $\SI{1500}{V}$. At an applied voltage of $\SI{2000}{V}$ the current initially stabilized at $\SI{8}{nA}$. After another minute at the same voltage, the current increased to $\SI{11}{nA}$ and a discharge channel formed between the electrodes (see figure \ref{fig:TMBi_discharge02}, lower panel). 
The current then increased from $\SI{20}{nA}$ to $\SI{23}{nA}$ at an applied voltage of $\SI{2500}{V}$. Increasing the voltage to $\SI{3000}{V}$ raised the current up to $\SI{41}{nA}$. At $\SI{3500}{V}$, the surface of the liquid started to move and at $\SI{4000}{V}$ the current increased up to $\SI{182}{nA}$ where the experiment was aborted (see figure \ref{fig:TMBi_exp2}). 
\subsection{Experiment 3: low pressure (<~40~mbar) and oxygen contamination}
\begin{figure}[h]
 \centering
 \includegraphics[width=1\columnwidth]{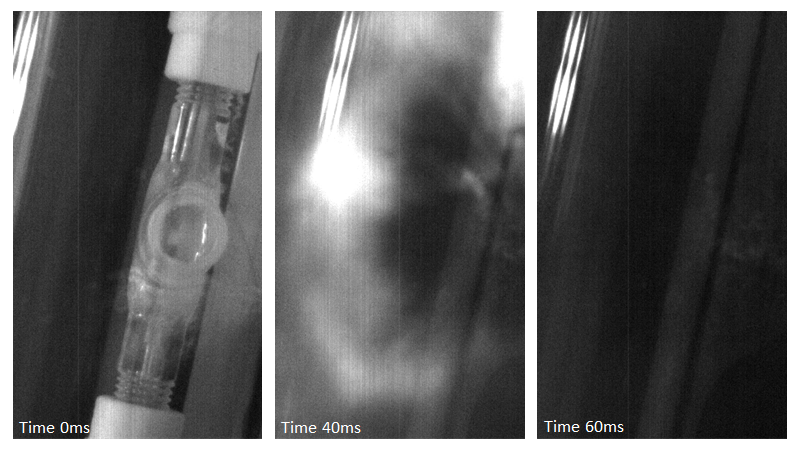}
 \caption{Photos of the sample cell before (left), during (middle) and after (right) the TMBi combustion. The milky discoloration of the TMBi described in the text is visible on the left photo. The combustion completely blackened the outer cylinder (right photo).}

 \label{fig:TMBi_explosion}
\end{figure}

In the third test, the pressure inside the cylinder was lowered to $<\SI{40}{mbar}$ by continuous pumping while applying a voltage of $\SI{-4000}{V}$. Due to the leak at the HV cable feedthrough, additional oxygen entered into the cylinder and reacted with the TMBi causing a milky discoloration (see figure~\ref{fig:TMBi_explosion}, left photo). 
Within the first 2~minutes, the current increased from \SI{82}{nA} to \SI{100}{nA}.
After 3~minutes of pumping the TMBi started to boil. The current increased to $\SI{3.8}{\micro\A}$ followed by an explosion of the TMBI which decomposed completely in the process (see figure~\ref{fig:TMBi_explosion}, middle and right photographs). 
\subsection{Experiment 4: low pressure argon atmosphere}
For the second round of measurements a new sample vessel was constructed, where the TMBi inlet is not positioned directly above the electrode gap, but on one side and at an angle. This allows to perform tests with an upright geometry where one electrode is submerged in the liquid and the other electrode positioned in the gas phase (see figure~\ref{fig:TMBi_exp5}, right photograph). An additional bulge above the electrode gap allows to observe the amount of gaseous TMBi formed in tests where the vessel is operated horizontally. As mentioned above, the leak observed in the previous measurements was fixed and, as an additional safeguard, the cylinder was placed in an argon filled containment to ensure that no oxygen can enter the sample vessel.

At the beginning of the measurement, the electrodes were completely covered with TMBi. The water jet pump was operated for 100~seconds reaching a pressure $\leq\SI{100}{mbar}$. Subsequently, the voltage was increased in steps of $\SI{500}{V}$ in 1~minute intervals up to a maximum of $\SI{4000}{V}$. Except for a temporary increase to \SI{0.6}{nA} at \SI{4000}{V} the measured current remained at \SI{0}{nA}.

\begin{figure}[h]
 \centering
 \includegraphics[width=1\columnwidth]{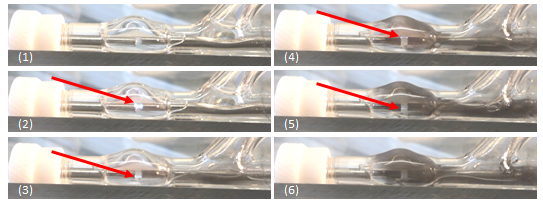}
 \caption{Experiment 4: TMBi in argon atmosphere at vapor pressures down to $\SI{20}{mbar}$. At $\SI{4000}{V}$, electrical breakdowns occurred between the electrodes leading to a decomposition of TMBi. The images were recorded in time from top left to bottom right over a one minute interval.}
 \label{fig:TMBi_breakdown_vaporpressure}
\end{figure}

To further lower the pressure, the water jet pump was operated for an additional 115~seconds and the TMBi filling level decreased due to evaporation until the electrodes were no longer completely covered (see figure \ref{fig:TMBi_breakdown_vaporpressure}, first photo). 
The voltage, which was turned off before pumping, was increased again step by step. At $\SI{3000}{V}$ the current increased to $\SI{0.2}{nA}$ for a few seconds, decreased again and remained at $\SI{0}{nA}$ for the rest of the measurement.

Finally, the water jet pump was operated permanently while keeping the voltage at $\SI{4000}{V}$. After ca. 2~minutes the liquid started to boil and electrical flash-overs occurred in the gas phase above the liquid. Parts of the TMBi decomposed and formed black residues on the glass surfaces of the sample container (see photographs 4-6 in figure \ref{fig:TMBi_breakdown_vaporpressure}). However, due to the absence of oxygen, the liquid did not spontaneously explode.
\subsection{Experiment 5: 1 bar argon atmosphere}

\begin{figure}[t]
 \centering
 \includegraphics[width=1\columnwidth]{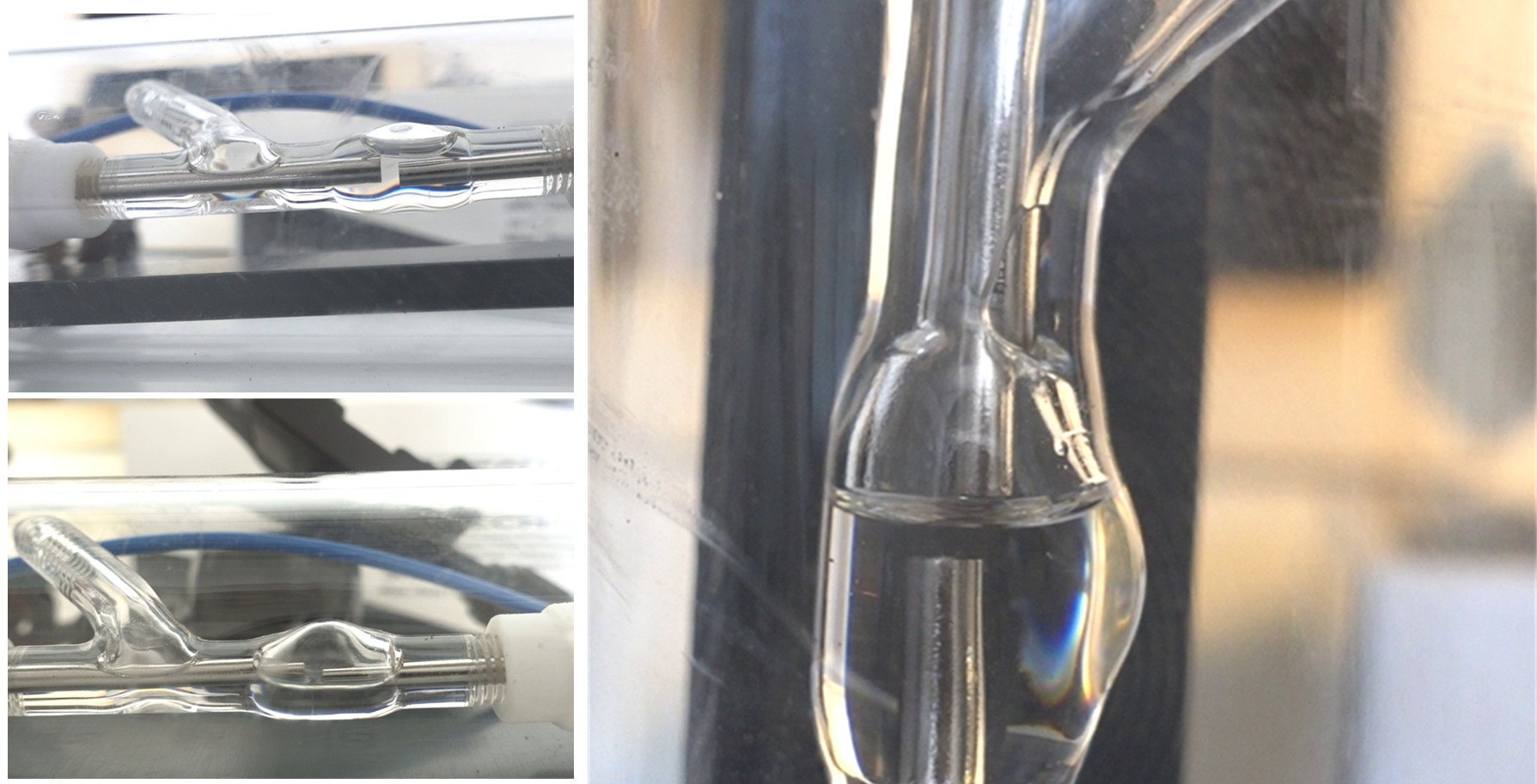}
 \caption{Experiment 5: Upper left: setup with electrodes completely covered with liquid. Lower left: reduced filling level with electrodes half covered with TMBi. Right: rotated setup with the lower electrode submerged in TMBi and the upper electrode in the gas phase. An argon pressure of 1~bar was present throughout the tests.}
 \label{fig:TMBi_exp5}
\end{figure}

The last series of measurements was again performed with different TMBi levels in the sample vessel but this time keeping a $\SI{1}{bar}$ argon atmosphere in the cylinder at all times. Tests were performed with completely covered electrodes, with partially covered electrodes and with the container in a vertical orientation, where the anode was submerged in TMBi while the cathode was positioned in the gas phase (see figure~\ref{fig:TMBi_exp5}). 
The current was measured while applying voltages up to $\SI{4000}{V}$. Besides the current, the liquid was monitored for fluid movements, the formation of bubbles, discharge channels or other disturbances. Neither during the initial ramping nor during prolonged measurements exceeding 30~minutes at \SI{4000}{V}, any changes were observed in the liquid or electric current.
Experiments with a 1~bar argon atmosphere were repeated several times with similar findings and without any breakdown occurrences, giving us confidence that this mode of operation indeed allows for stable operation of TMBi filled measurement cells.

\label{sec:exper}

\section{Conclusion}
The objective of this study was to investigate the behaviour of TMBi under different ambient conditions and different electric field strengths to find conditions for stable and safe operation of a liquid-based ionization detector.
For that purpose a test setup containing a TMBi filled sample cell was constructed 
that allowed to work under an argon atmosphere at different pressures while applying varying electric fields to the liquid. During the first measurements an oxygen contamination was present inside the test cylinder due to a leaky feedthrough. This actually allowed us to study the effect of an electro-negative contamination in the liquid on the high-voltage stability.

Our findings may be summarized as follows: Under argon atmospheres at 1~bar pressure we did not observe breakdowns of the electrical potential in our measurement cell up to maximum field strengths of $\SI{20}{kV/cm}$. The presence of oxygen contaminations in the first series of measurements did, however, led to fluctuating currents in the nA range, which were not observed under a completely pure argon atmosphere.

Lowering the pressure to about 40~mbar, corresponding to the TMBi vapor pressure at room temperature, lead to the formation of a discharge channel in the liquid and a steady increase in the current in the presence of oxygen contaminations (the measurement was stopped at a current of 182~nA). Further reduction of pressure caused the liquid to boil and, in the presence of oxygen, led to a spontaneous combustion of the TMBi.

With the improved setup, allowing measurements under a pure argon atmosphere, no continuous current was observed even when the electrodes were not completely covered in TMBi. At pressures below the boiling point the TMBi started to decompose at the applied field of $\SI{20}{kV/cm}$, but without the explosive character observed earlier. The presence of 1~bar argon prevented the liquid from boiling at room temperature. A slight overpressure compared to the ambient air pressure also minimizes the instrument's risk of oxygen ingress. 
Although an even higher pressure would result in a higher breakdown voltage, it would probably be incompatible with many vacuum components.

If the electrodes are not completely immersed in the liquid, the breakdown mechanisms in gases will come into effect causing a reduction of the allowable field strength (see Paschen's curve in figure~\ref{fig:paschen06}).
Also, our tests show that electro-negative contaminations in the liquid have an important and detrimental effect on the electric stability of the medium.

When going to pressures close to the vapor pressure of the liquid, small additional pressure or temperature changes will cause the liquid to boil and will change the conditions for an electrical breakdown in the medium. In a liquid argon study, the breakdown voltage decreased from $>\SI{100}{kV/cm}$ to $\SI{40}{kV/cm}$ when the liquid started to boil~\cite{bay2014evidence}. Thus, special care has to be taken to avoid such situations.

We conclude that in the presence of a pure argon atmosphere at 1~bar pressure, a detector with plane-parallel electrodes completely covered in a purified dielectric liquid such as TMBi remains stable against electrical breakdown at least up to electric field strengths of $\SI{20}{kV/cm}$, which we had available during our measurements.
\label{sec:disc}

\acknowledgments

We would like to thank Deutsche Forschungsgemeinschaft (DFG), project numbers WE 1843/8-1 and SCHA 1447/3-1, French national research agency (ANR), project number ANR-18-CE92-0012-01, and Axel Buß for providing the photography in figure \ref{fig:Tube}.

%\paragraph{Note added.} This is also a good position for notes added
%after the paper has been written.

\vspace{-0.1cm}
\bibliographystyle{JHEP}
\bibliography{99_bibliothek}

\end{document}